\documentclass[preprint2]{emulateapj}
\usepackage{subfigure}
\usepackage{rotating}

\slugcomment{accepted by AJ}

\shorttitle{NGC~6441 RR Lyrae stars}
\shortauthors{Kunder et al.}

\begin{document}

\title{Radial velocities of RR Lyrae stars in and around NGC 6441}


\author{Andrea Kunder\altaffilmark{1},
Arthur Mills\altaffilmark{1},
Joseph Edgecomb\altaffilmark{1},
Mathew Thomas\altaffilmark{1},
Levi Schilter\altaffilmark{1},
Craig Boyle\altaffilmark{1},
Stephen Parker\altaffilmark{1}, 
Gordon Bellevue\altaffilmark{1},
R. Michael Rich\altaffilmark{2},
Andreas Koch\altaffilmark{3},
Christian I. Johnson\altaffilmark{4}
David M. Nataf\altaffilmark{5}
}
\altaffiltext{1}{Saint Martin's University, 5000 Abbey Way SE, Lacey, WA, 98503} 
\altaffiltext{2}{Department of Physics and Astronomy, University of California at Los Angeles, Los Angeles, CA 90095-1562, USA}
\altaffiltext{3}{Astronomisches Rechen-Institut, Zentrum f\"ur Astronomie der Universit\"at Heidelberg, 
M\"onchhofstr.\ 12--14, 69120 Heidelberg, Germany}
\altaffiltext{4}{Harvard-Smithsonian Center for Astrophysics, Cambridge, MA 02138}
\altaffiltext{5}{Center for Astrophysical Sciences and Department of Physics and Astronomy, The Johns Hopkins University, Baltimore, MD 21218, USA}

\begin{abstract}
Detailed elemental abundance patterns of metal-poor ($\rm [Fe/H]$$\sim$$-$1~dex) stars
in the Galactic bulge indicate that a number of them are consistent with globular cluster (GC) stars 
and may be former members of dissolved GCs.  This would indicate that a few per cent of the Galactic
bulge was built up from destruction and/or evaporation of globular clusters.
Here an attempt is made to identify such presumptive stripped stars originating from the massive, inner 
Galaxy globular cluster NGC~6441 using its rich RR Lyrae variable star (RRL) population.  
We present radial velocities of forty RRLs centered on the globular cluster NGC~6441.
All of the 13 RRLs observed within the cluster tidal radius have velocities consistent with
cluster membership, with an average radial velocity of 24 $\pm$ 5~km~s$^{-1}$ and a 
star-to-star scatter of 11~km~s$^{-1}$.  This includes two new RRLs that were previously
not associated with the cluster.  Eight RRLs with radial velocities consistent
with cluster membership but up to three time the distance from the tidal radius are also
reported.  These potential extra-tidal RRLs also have exceptionally long periods,
which is a curious characteristic of the NGC~6441 RRL population that hosts 
RRLs with periods longer than seen anywhere else in the Milky Way.
As expected of stripped cluster stars, most are inline with the cluster's orbit.
Therefore, either the tidal radius of NGC~6441 is underestimated and/or we
are seeing dissolving cluster stars stemming from NGC~6441 that are 
building up the old spheroidal bulge.
\end{abstract}


\section{Introduction}
Globular clusters (GCs) are relics from the early universe, and as such, play an important role
in deciphering the early history of the Galaxy.  Their interplay within the Galactic bulge, 
however, is poorly understood, due in part to several strange bulge globular clusters 
which have properties not seen anywhere else in the Milky Way.
For example, Terzan~5 is a globular cluster harboring stars of hugely different ages, 
implying it was previously a massive structure formed at the epoch of the Milky Way 
bulge formation \citep{ferraro16}.
Like Terzan~5, two further GCs in bulge, NGC~6569 and NGC~6440, have double
horizontal branches \citep{mauro13}, a morphology not seen anywhere else in the Milky Way, and
suggestive that there are more of these ancient fossil remnants of massive clumps lurking in the bulge.

NGC~6441 is also one of these curious inner Galaxy GCs.
It has a very extended blue horizontal branch \citep{rich97} in spite of
its high metallicity of $\rm [Fe/H]\sim-0.41 \pm$ 0.06,  rms = 0.36~dex, on the
Carretta \& Gratton metallicity scales \citep{gratton07, clementini05}, and 
traditional evolutionary scenarios can not account for this property following the
discovery that their RR Lyrae variables (RRLs) have exceptionally long periods \citep{layden99, pritzl01}.
The rich population of RRLs in NGC~6441 \citep[more than 50,][]{clement01}, 
is further curious given the paucity of RRLs in metal-rich systems.  For example, less 
than 10\% of field RRLs within 2 kpc of the Sun have $\rm [Fe/H] > -$0.5 dex, and this is true 
also of the preponderance of RRLs within Galactic GCs.  

The mix of at least two stellar populations observed in NGC~6441 is thought to be composed of 
stars with distinct values of helium abundances \citep{gratton07, bellini13}.  
One formation scenario that explains why this globular cluster 
hosts different stellar populations is that massive and complex clusters such as NGC~6441
may have been former dwarf galaxy cores -- remnants of a larger system, such as a 
dwarf galaxy or star cluster complex that merged with the Milky Way and is currently being 
pulled apart by the tidal forces of the Milky Way \citep{bekki16, bekki17}.  
This would be consistent with its large mass -- fifth in rank after clusters 
$\omega$~Cen, M~54 at the center of Sagittarius dwarf galaxy, NGC~6388 and NGC~2419.  
This would also be consistent with its extended horizontal branch (EHB), as several studies have
argued that EHB clusters are of extragalactic origin, formed probably as nuclei
in dwarf satellites that were accreted by the Milky Way \citep{lee07, bekki06}.
Further evidence that this globular cluster was accreted is that 
its RRLs have periods that are longer than seen in other Milky Way globular 
clusters \citep{pritzl00, pritzl01}.  

Because NGC~6441 lies in a field that is very crowded, severe contamination by foreground 
(mainly bulge) field stars prevents the clear extent of this GC.  Its position in the Galaxy is 
also where the reddening is severe, not only along its line of sight where $E(B-V)$ is 0.47 mag 
\citep[from][2010 edition]{harris96}, but it also has extreme differential reddening 
throughout the cluster with $\Delta$$E(B-V)\sim$0.14 mag \citep{law03}. 
Therefore, identifying possible tidal tails around NGC~6441 is a 
challenging task.  

It is especially intriguing to search for tidally stripped stars, however, since recent results
indicate their are field stars in the inner Galaxy with abundances 
that are typically found in GC stars \citep{fernandeztrincado17, schiavon17a}.  These
stars appear to be homogeneously distributed with a 
metallicity distribution peaking at $\rm [Fe/H]$$\sim$$-$1.  The current
best interpretation is that they are members of dissolved GCs \citep{schiavon17a, 
schiavon17b, fernandeztrincado17}.  Since the elemental abundances of these
potential GC stripped stars coincide with stars formed in the later stellar generations of 
the GC, the RRLs in and around NGC~6441, which are also thought to be a third (or later)
generation of cluster stars \citep{lee14}, are particularly good probes for signatures of 
dissolving GC stars.

From the orbit of NGC~6441, it is known that although
often labeled as a bulge cluster \citep[e.g.,][]{origlia08, bellini13}, 
and although indeed confined to the inner $\sim$3-4 kpc of the 
Galaxy, its velocity components are not consistent with it being a member of the 
central bulge/bar \citep{casetti10, bica15}.
Instead this cluster, along with the globular cluster NGC~6388, can be thought of as 
members of a pressure supported-spheroidal bulge with low rotation and high velocity 
dispersion, and therefore would have a different origin than the bulge/bar.  
Especially in light of the recent finding that
the inner Galaxy RR Lyrae stars are also a population of stars consistent with a spheroidal 
component that formed before the bulge/bar 
that dominates the mass of the bulge \citep{kunder16}, it is important to search for and understand
other old systems that may be part of the ancient inner Galaxy.  


In this paper, we report on the kinematics of 40 RR Lyrae stars centered on NGC~6441, but extending
spatially out to more than half a degree (4 times that of the cluster tidal radius).
We have taken advantage of the pulsation properties of RRLs to further establish membership of these stars
and investigate the connection between the NGC~6441 globular cluster, the underlying RRL field 
star properties and the build-up of the bulge/bar.  Lastly, an orbital integration of the
cluster is carried out to determine its orbit around the Milky Way and hence the location in the
sky where stripped tidal debris is most likely to reside.



\section{Observations and Radial Velocity}
\begin{figure*}
\centering
\mbox{\subfigure{\includegraphics[height=8.4cm]{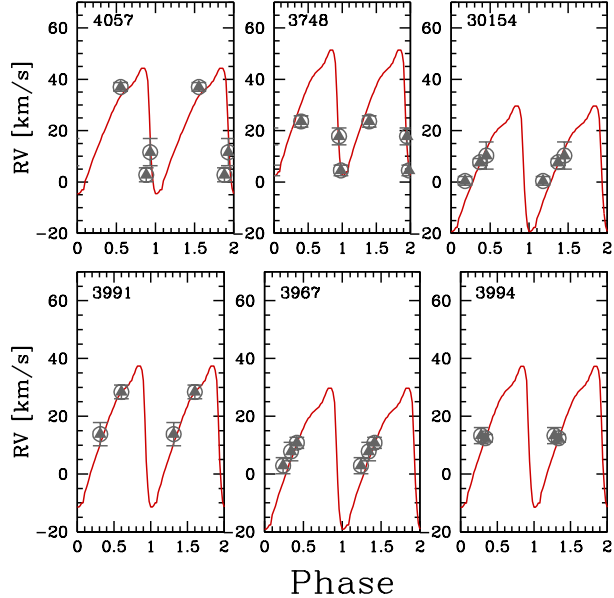}}\quad
} 
\caption { The line-of-sight radial velocity versus pulsational phase for typical 
stars in our sample, over-plotted on a
fundamental mode RRL radial velocity template and scaled by its $V$-amplitude \citep{liu91}.
}
\label{craig}
\end{figure*}

Observations were carried out using the AAOmega multi-fiber spectrograph on 
the Anglo-Australian Telescope (AAT) on three separate nights:
09 August 2016, 10 August 2016, and 18 June 2017
(OPTICON proposal code 17A/051, NOAO PropID: 2017A-195 and NOAO PropID: 2016B-0058 PI: A. Kunder).  The spectrograph was centered on 8600\AA, with the 580V and 1700D gratings to probe the Calcium Triplet,
so the wavelength regime of $\sim$ 8300\AA~to 8800\AA~was covered at a resolution of R$\sim$10,000.
The exposure times ranged from 1800 to 7200 seconds, and one exposure per night was taken,
meaning we had 3 epochs for each RRL.

Targets were selected using the Optical Gravitational Lensing Experiment (OGLE) 
catalogue of RRLs \citep{pietrukowicz12} and
we prioritized those stars with existing high-resolution spectroscopy from \citet{clementini05} to
test the CaT metallicity relation.  Unfortunately the poor weather conditions did not allow for
high enough signal-to-noise (SNR) ratios for robust equivalent width determinations and hence 
metalliciites for the observed stars.  The spectra had SNR ratios ranging from 7 to 30.  

An identical method to determine radial velocities as presented in \citet{kunder16} was 
carried out.  Briefly, the observations were first phased by the stars known OGLE period.  Then the 
radial velocity template from \citet{liu91} was used to find the center-of-mass radial velocity for 
each star, upon being scaled to the correct amplitude for each star.
Again, the zero-point in phase was fixed using the time of maximum 
brightness as reported by OGLE-IV \citep{soszynski14} and we verified that all three
observations fit the shifted pulsation curve.  An example of a typical radial velocity curve presented here,
as well as the photometric light curve from OGLE data, is shown in Figure~\ref{craig}.

The only difference employed here from \citet{kunder16} is that the zero-point in radial velocity 
was calculated by finding the observed center of all three Calcium triplet lines and comparing those
to the known Calcium triplet lines (8498.03 \AA, 8542.09 \AA, 8662.14 \AA).  In contrast,
\citet{kunder16} cross-correlates a spectra of a giant star (with a well-determined radial velocity) to each
RR Lyrae star, and the peak of the cross-correlation peak is what was adopted as the radial velocity measurement.
We believe that although more time-consuming, directly measuring the center of the Calcium triplet lines 
leads to more accurate zero-point determination.  This is because the
shape and features in the RR Lyrae spectra change over it's $\sim$0.5 day pulsation cycle due to 
the temperature variations of the star, so using just one template to cross-correlate all the
RR Lyrae spectra could give rise to uncertainties for pulsation phases that especially are hotter 
than that of the template, where e.g., more Hydrogen Paschen lines are present.

Table~1 gives the OGLE-ID (1), the Galactic $l$ (2) and Galactic $b$ (3) as provided by OGLE, 
the star's time-average velocity (4), the number of epochs used for the star's time-average velocity (5), 
the $I$-band amplitude (6), the $I$-band magnitude (7),
the $V$-band magnitude (8) and the period of the star (9) as calculated by OGLE, 
the distance from the cluster center in arc minutes (10), and lastly any notes (11).  In particular,
the notes identify the RRLs observed here with known NGC~6441 RRLs listed in \citet{clement01}.

\section{Results}
\subsection{Cluster membership}
\begin{figure*}
\centering
\mbox{\subfigure{\includegraphics[height=8.4cm]{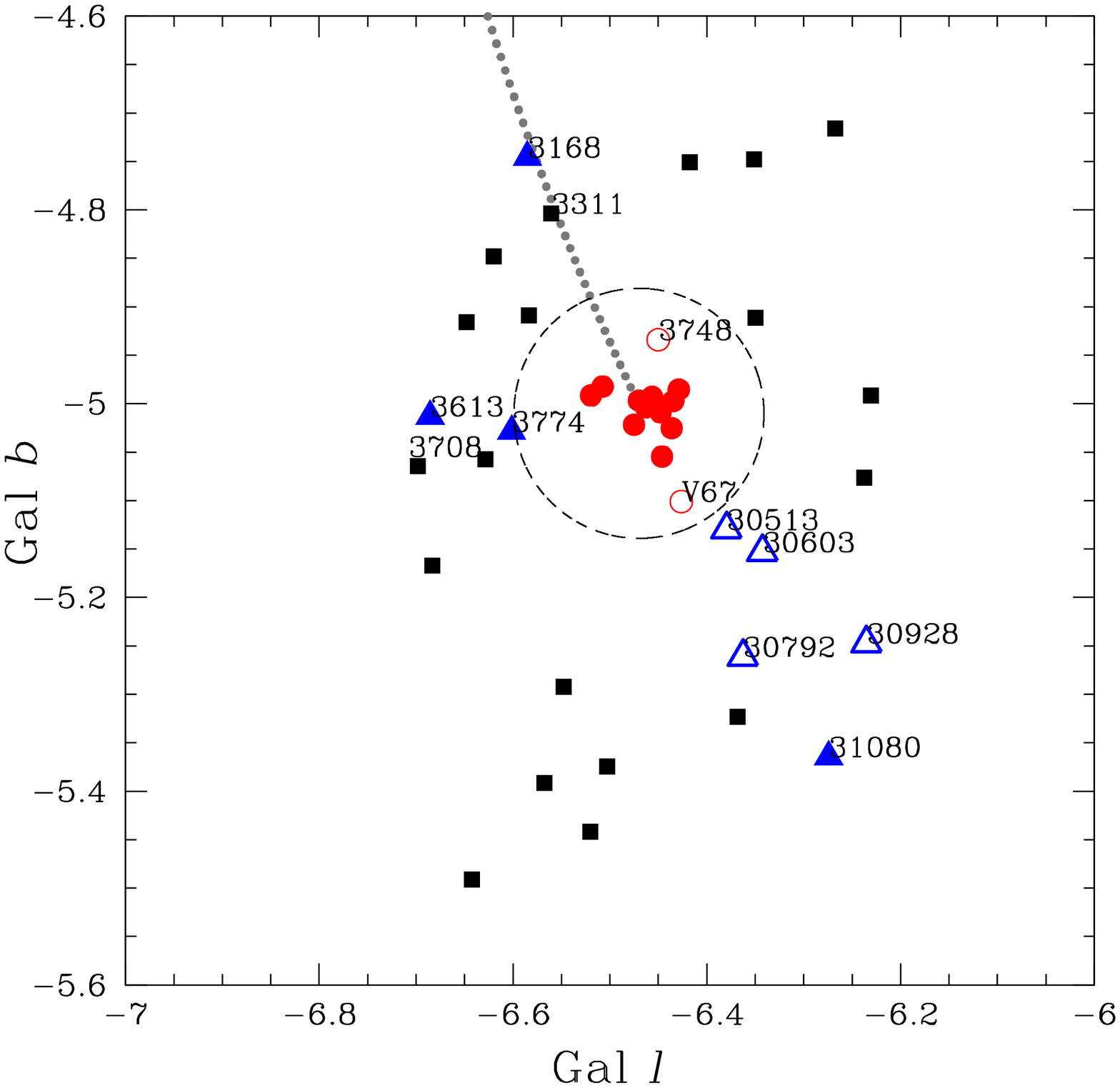}}\quad
\subfigure{\includegraphics[height=8.4cm]{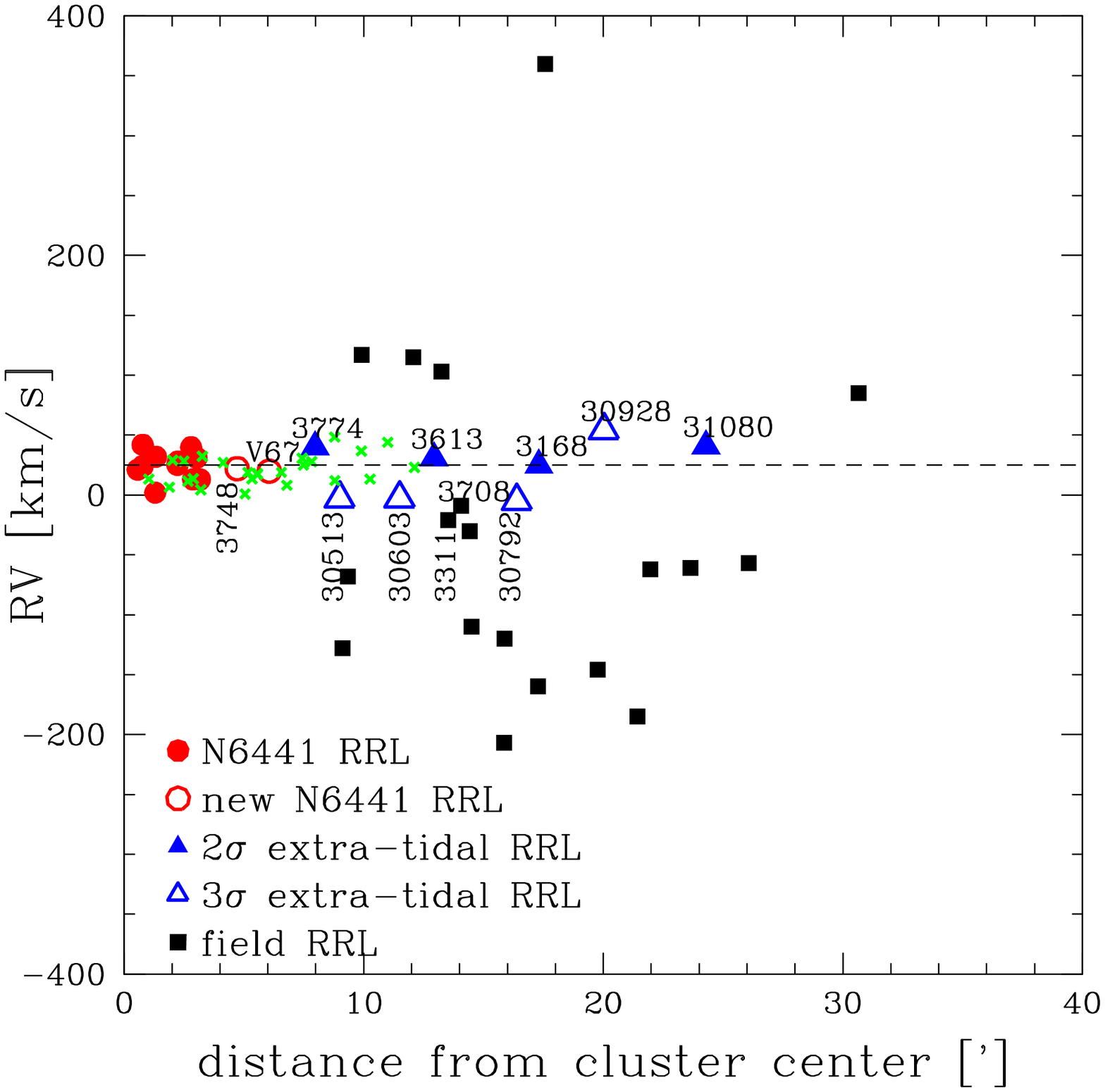} }
} 
\caption { {\it Left:} The spatial position of the RRLs in our sample, with the cluster tidal radius
over-plotted.  The filled red circles designate known cluster members, and the open circles
designate the two RRLs within the tidal radius of NGC~6441 that have velocities consisted with
cluster membership but were not previously associated with the cluster.  
The blue triangles
indicate stars outside the tidal radius but with radial velocities consistent with cluster membership.
{\sl Right:} The radial distribution of radial velocities in our sample.  Green crosses indicate
the cluster giants from \citet{gratton07}.  
}
\label{levi_mathew}
\end{figure*}

The spatial positions of the observed stars are shown in Figure~\ref{levi_mathew} (left panel), and the
tidal radius of $\sim$7.7 arcmin \citep[0.129$^\circ$, 2010 edition of][]{harris96} is shown.  None of our 
observed RRL fall within the core radius or half-light radius.  As expected,
the density of the observed RRLs is highest closest to the center of the cluster, and decreases
as the tidal radius of the cluster is approached.  All RRLs observed within the tidal
radius have velocities consistent with cluster membership.  This is true even for V67, 
which was speculated to
be a field star by \citet{pritzl01} because of its large distance from the cluster center as well as
its uncertain/blended photometry.  We have also observed
an RRL within the tidal radius of NGC~6441 that is not included in the \citet{clement01} catalog 
of variable stars, but that has a velocity consistent with cluster membership (OGLE-03748).


Also shown in Figure~\ref{levi_mathew} (right panel) is the radial velocity of the stars as a function 
of the distance from the center of the cluster.  The group of thirteen stars located within
the cluster tidal radius (red circles) show similar kinematic properties, 
clustering around a mean velocity of 24 $\pm$ 5~km~s$^{-1}$, with a star-to-star scatter of 
11~km~s$^{-1}$.  This radial velocity is in excellent agreement found from a sample of 
25 NGC~6441 giants which were vetted not only by radial velocity but also by 
abundance \citep{gratton07}.  In particular, \citet{gratton07} report an average radial 
velocity of 21 $\pm$ 2.5 km~s$^{-1}$, with a star-to-star scatter of 12.5 km~s$^{-1}$.
From spectroscopy of 12 RRLs in NGC~6441, \citet{clementini05} find  
an average radial velocity of $-$1 $\pm$ 7 km~s$^{-1}$, with a star-to-star scatter of 12 km~s$^{-1}$.
Although the scatter is similar to what is found here, the average velocity reported by
\citet{clementini05} is considerably lower.  However, it is worth noting that \citet{clementini05} 
measured one epoch per RRL, and so their mean 
radial velocity values included phase-dependent contributions due to the RRL pulsations.

As with the giant stars observed by \citet{gratton07}, we find RRLs that have radial velocities 
consistent with cluster membership but farther than the nominal tidal radius of 0.129 degrees.
In fact, eight RRLs outside the tidal radius have radial velocities within 3-sigma of that of the cluster.
Four of these (blue closed triangles in Figure~2) have velocities only $\sim$2$\sigma$ from 
the cluster mean, and four (open triangles) have velocities within $\sim$3$\sigma$ of
the cluster mean.  We label the stars having a radial velocity within $3\sigma$ of
the cluster mean as candidate cluster members.
Given the large extent of the NGC~6441, and also that radial trends in
radial velocity across clusters are expected \citep[e.g.,][]{perryman98}, at least some
of these stars likely have some connection with the cluster.


This is strengthened because they are spatially located along the orbital path of the cluster
(see \S\ref{orbit}) and as shown in the next section, these stars have especially long 
periods for their amplitudes, in agreement with the extreme NGC~6441 RRL pulsation properties.

\subsection{Period-amplitude diagram and color-magnitude diagram}
\begin{figure*}
\centering
\mbox{\subfigure{\includegraphics[height=7.4cm]{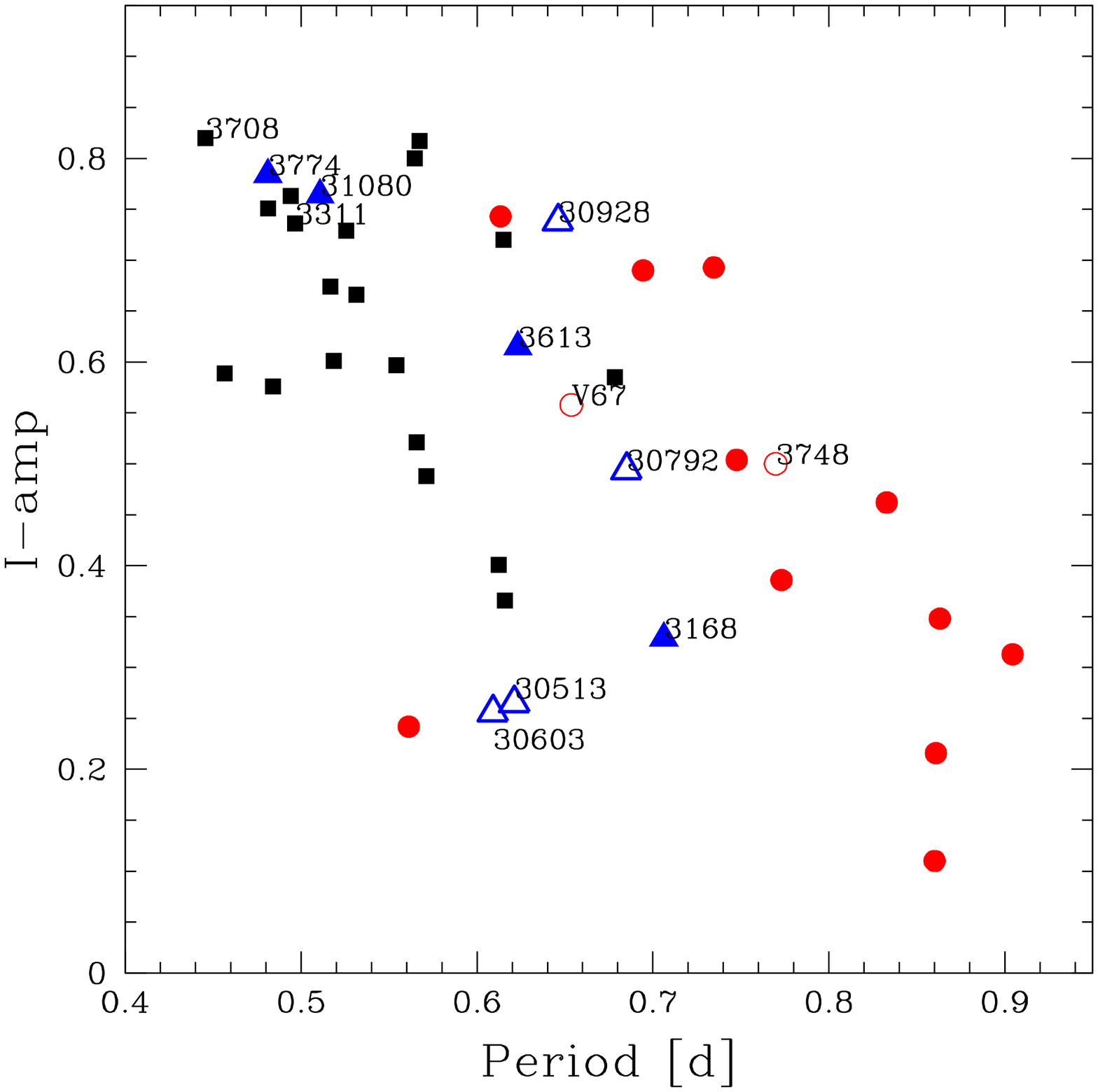}}\quad
\subfigure{\includegraphics[height=7.4cm]{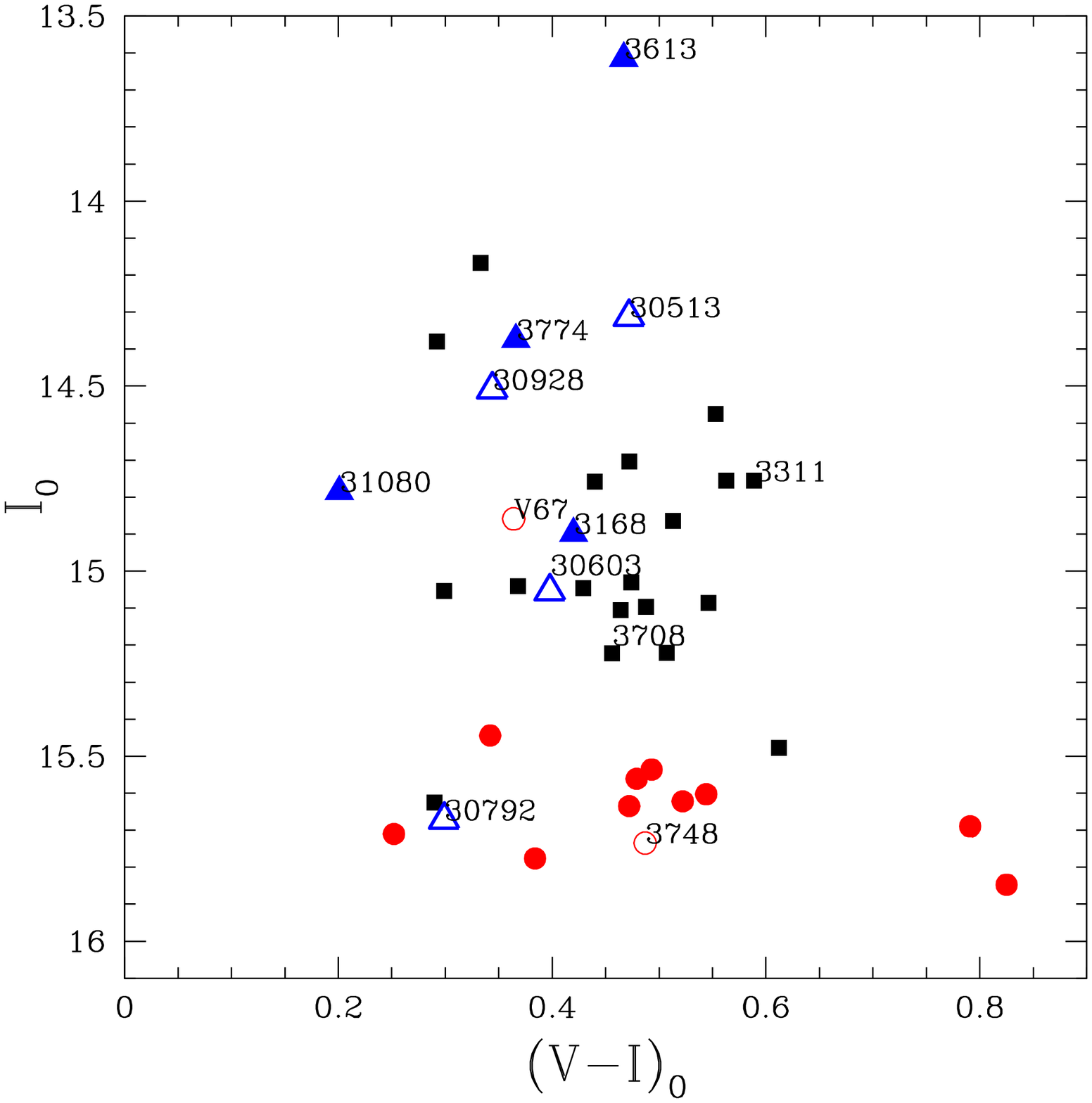} }
} 
\caption { The period-amplitude diagram (left) and color-magnitude diagram (right) for the RR Lyrae stars in and around NGC~6441.
}
\label{arthur}
\end{figure*}

Figure~\ref{arthur} (left panel) shows the periods and amplitudes of our observed stars as well
as the colors and magnitudes of the observed RRLs (right panel).  It is evident that
the NGC~6441 cluster members have considerably longer periods than those in the field.  
In fact, the NGC~6441 RRLs have periods that are so
unusual, that they have been given their own Oosterhoff group \citep[OoIII --][]{pritzl03, jang15}.
One popular explanation of their long periods is that these RRLs have Helium enhancements of 
$Y \sim$0.35 - 0.4 dex \citep[e.g.,][]{caloi07, busso07, tailo17}.  
The potential extra-tidal RRLs have periods that are also separated from the typical bulge field 
population, in a sense that they are shifted to longer values.  

\begin{figure*}
\centering
\mbox{\subfigure{\includegraphics[height=7.4cm]{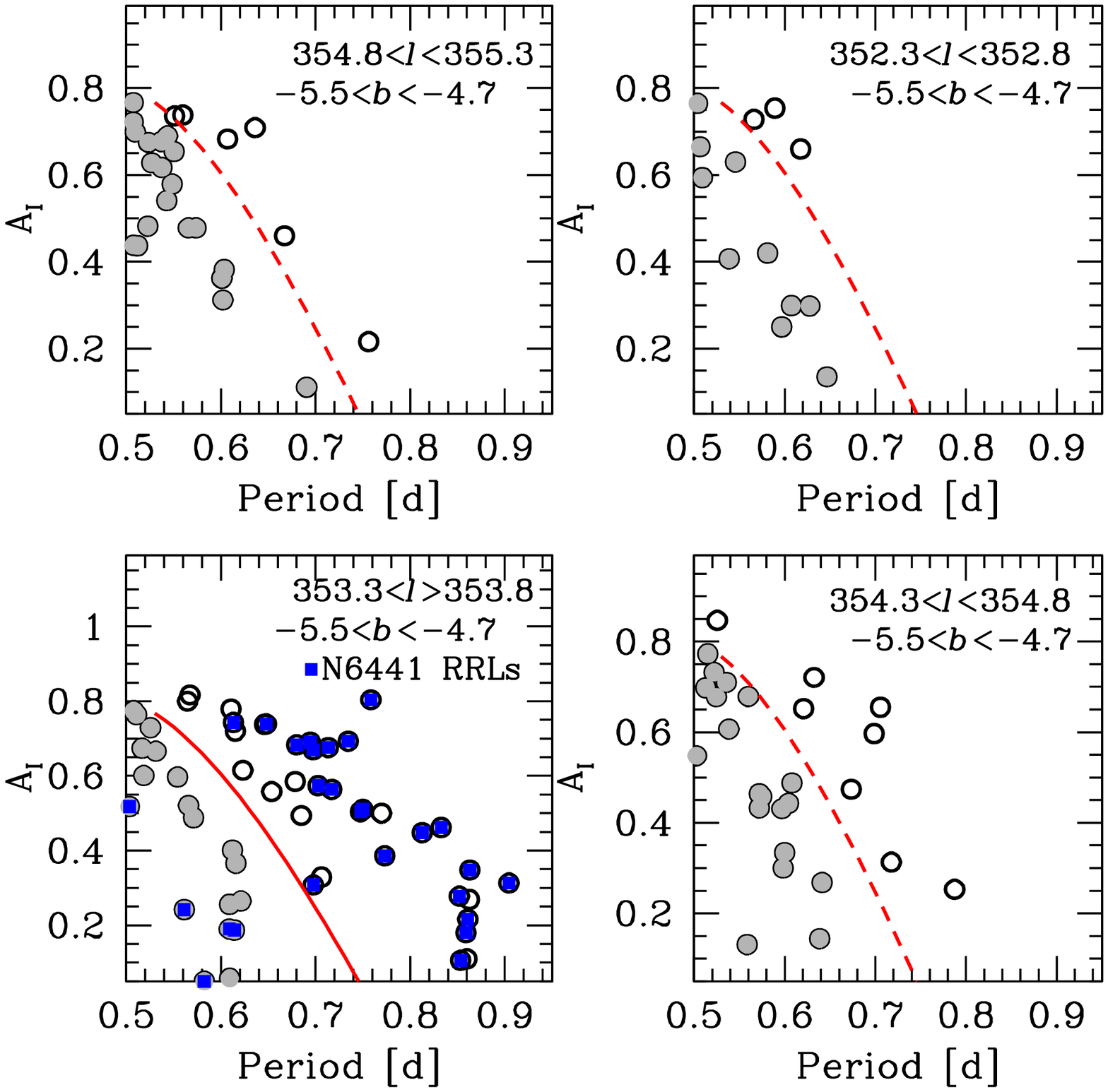}}\quad
\subfigure{\includegraphics[height=7.4cm]{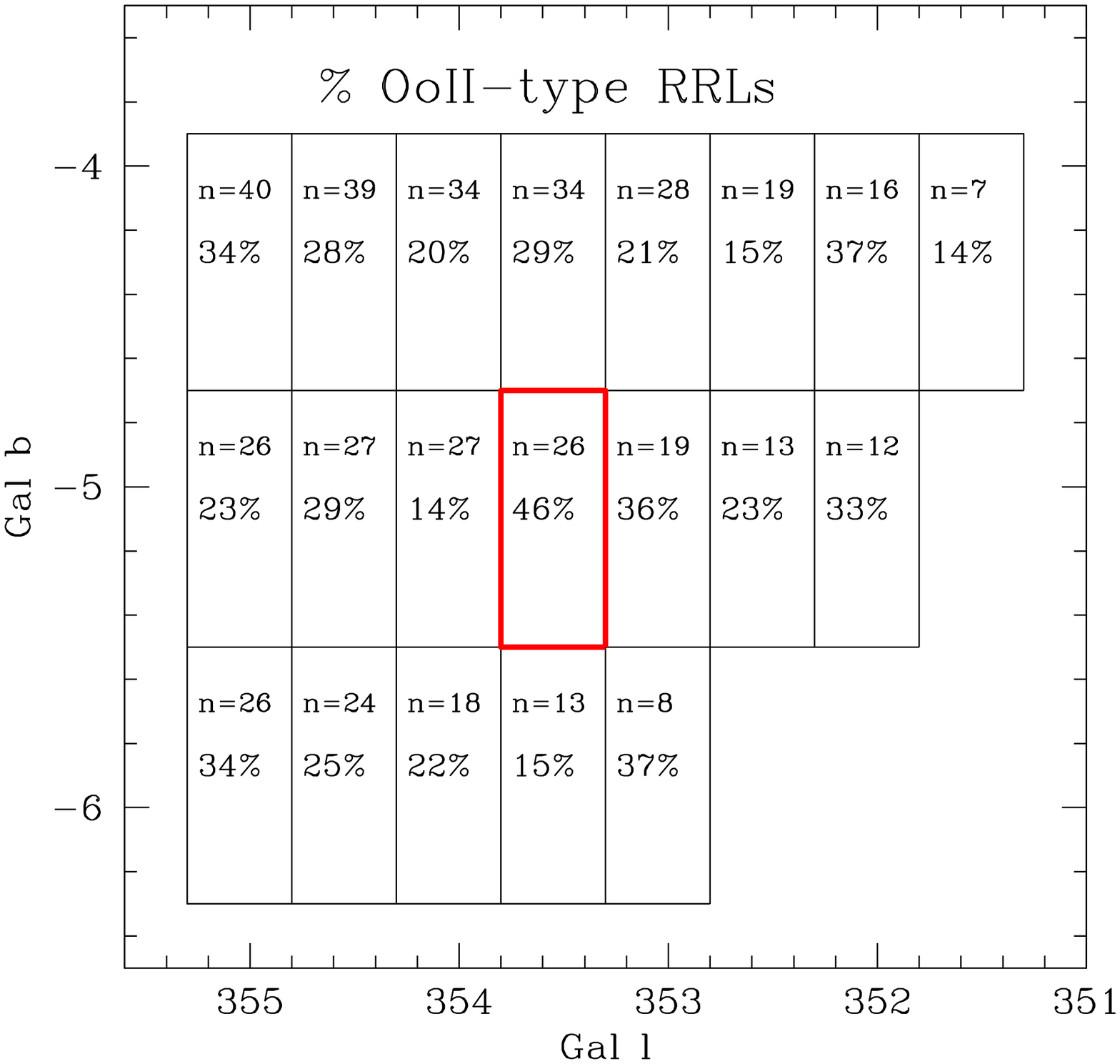} }
} 
\caption { {\it Left:} The period-amplitude diagrams of RRLs in our observed field (bottom-left) compared to 
fields close to but not overlapping our field.  The field RRLs surrounding NGC~6441 have a larger
preponderance of long period (OoII-type) RRLs than field RRLs 0.5 to 2 degrees further from the
cluster.  {\bf The stars with OoII-type properties are indicated by the open circles. }
{\it Right:}  A spatial map indicating the percent of OGLE OoII-type RRLs with periods greater than 
0.5 d for a given 0.5$^\circ$ x 0.7$^\circ$ field.  Our field observed here is marked by the bold (red) line.  
The total number of RRLs with periods greater than 0.5 d is also indicated.  
}
\label{referee}
\end{figure*}

It is interesting to assess how likely it is for random field stars to fall in different parts of the 
period-amplitude diagram.  
Because the OoI- and OoII-type RRLs are very difficult to disentangle at periods less than 0.5 days,
and because here we are interested in the preponderance of OoII- (or OoIII-type) stars, 
we focus on stars only with periods greater than 0.5 days.
Figure~\ref{referee} shows the OGLE stars in our observed 
0.5$^\circ$ x 0.7$^\circ$ field with periods greater than 0.5~d, compared to the OGLE stars in 
neighboring fields with periods larger than 0.5~d.  We designate 
an OoII-type RRL as those with 
\begin{equation}
A_I > -1.642-13.78 (log(P)-0.03) - 19.298 (log(P)-0.03)^2
\end{equation}
where $A_I$ is a star's I-amplitude and $P$ is the star's period \citep[see][]{cacciari05}.
It is immediately apparent that for stars with $P>$ 0.5~d, NGC~6441 has a larger percentage 
of OoII-type stars than the field ($\sim$81 \%).  In contrast, $\sim$46\% of the RRLs surrounding 
this cluster with $P>$ 0.5~d are OoII-type RRLs.
For fields located $\sim$0.5 degrees away from the cluster, 
an even smaller percentage, between 23 - 30\% of stars, are OoII-type stars (right panel of Figure~\ref{referee}).  
Therefore, 
the field RRLs immediately surrounding NGC~6441 have a larger percent of OoII-type
RRLs than fields 0.5 to 2 degrees away from the cluster, suggestive that some of
our ``field" RRLs do belong to NGC~6441. 



Because of the unusual RRL properties of the NGC~6441 stars, their absolute magnitudes 
are thought to be brighter than normal, making it difficult to know what absolute magnitude 
to use for distance determinations \citep[this is true also for the red giant branch bump, as .
shown by e.g.,][]{nataf13a}.  
Varying and patchy reddening in the 0.5 degree field studies is also a cause for concern when 
interpreting the CMD.  Both the reddening maps from OGLE \citep{nataf13b} as well as those 
from minimum-light colors of the RRLs \citep{kunder10}, show that the RRLs surrounding 
NGC~6441 are more extincted than those within the  cluster tidal radius by $\sim$0.2~mag in $I$.

Finally, the differing periods of the central NGC~6441 RRLs with those on the outskirts
suggest that the stars within the tidal radius of NGC~6441 are $\sim$0.2 mag brighter 
than those in the outskirts.  A range of absolute magnitudes of RRLs in and around this 
cluster is also expected from $\rm [Fe/H]$ variations seen in the cluster RRLs \citep{clementini05}.
We therefore use the color-magnitude diagram as only a rough guide for evaluating cluster 
membership.

The known cluster members have magnitudes that make their identification with 
NGC~6441 straight-forward -- they are fainter and located at a distance of $\sim$11.6~kpc, 
whereas the distance of the bulge peaks at $\sim$8~kpc.
Indeed, it was photometrically that \citet{layden99} associated these RRLs with NGC~6441.
Only one extra-tidal candidates has a magnitude consistent with that of the cluster.
The majority of extra-tidal candidates have magnitudes that are $\sim$1 mag brighter than
expected -- they have magnitudes similar to V67, a star spectroscopically studied here with 
a velocity consistent with cluster membership, but initially thought to be too bright to be a cluster 
member \citep{pritzl01}.  Note that when using the {\it Hubble Space Telescope (HST)}, \citep{pritzl03} 
found that previous ground-based photometry of the NGC~6441 RRLs 
is largely affected by blends, resulting in photometry errors of up to 1.5 mag in the $V$ band.


%
%

\subsection{The field RRLs around NGC 6441}
\label{field}
The 27 field RRLs (including the extra-tidal candidates) have a mean heliocentric velocity of 
$-$15~km~s$^{-1}$ (galactocentric velocity of $-$32~km~s$^{-1}$) with a star-to-star 
scatter of 117~km~s$^{-1}$.  Removing the eight potential 
extra-tidal candidates, the field RRLs have a 
mean velocity of $-$30~km~s$^{-1}$ (galactocentric velocity of $-$48~km~s$^{-1}$) with a star-to-star scatter 
of 137~km~s$^{-1}$.  This large velocity dispersion is in stark contrast
to what is seen for the giants and red clump stars in the bulge that trace out the 
Galactic bulge/bar \citep[e.g.,][]{kunder12,ness13,zoccali14}.  
It is also in contrast to the velocity dispersion of the \citet{schiavon17a} sample of 59 N-rich stars 
which are those linked to GC stars.  Despite the fact that the N-rich stars in \citet{schiavon17a}
are located within $|b| < 5^\circ$, the location in the bulge where the largest dispersion in 
velocities is observed \citep[$\sim$100~km~s$^{-1}$, e.g.,][]{kunder12}, these
stars have a velocity dispersion of only 86 km~s$^{-1}$.  

Instead, the velocity dispersion is consistent with the 
hot population identified by \citet{kunder16} and also seen in targeted metal-poor star studies of
the bulge \citep[e.g.,][]{howes14, howes15, howes16, schlaufman14, koch16, garciaperez13, schultheis15}.
Note, however, that as \citet{walker91} show, the metallicity distribution of the bulge RRLs may not
necessarily be as low as the metallicities probed by the cited studies.  The bulge RRLs stand out as a
different population within the inner Galaxy because of their old age, predating the stars that dominate
the mass of the inner Galaxy.  It is likely that the similarities in radial velocities and dispersions of the 
extremely metal-poor stars being currently discovered means that these stars have similar ages as the
RRL population.

\begin{figure*}
\centering
\mbox{\subfigure{\includegraphics[height=8.4cm]{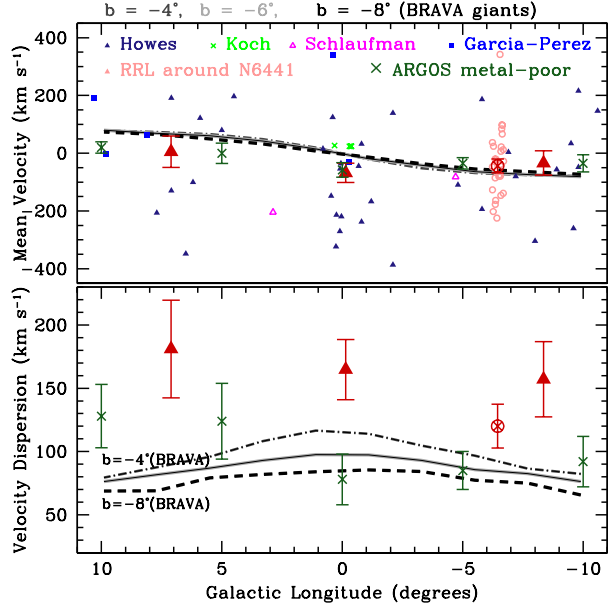}}\quad
} 
\caption { The velocity dispersion profile (bottom) and rotation curve (top) for the 
metal-poor stars ($\rm [Fe/H] < -$2.0) observed in the bulge compared to that of the field RRLs surrounding
NGC~6441 and the BRAVA giants at $b$ = $-$4$^\circ$, $-$6$^\circ$, 
and $-$8$^\circ$ strips \citep{kunder12}.  
The circled asterisks designates the mean velocity and mean velocity dispersion
of the RRLs presented here.
The individual metal-poor star measurements are given in the top panel, and the 
large red circles indicate the mean Galactocentric velocity (top) and velocity dispersion (bottom).
The $\sim$100 metal-poor ARGOS stars ($\rm [Fe/H] < -$1.0) 
are shown separately, as the individual ARGOS stars and stellar parameters are non-public.
The metal-poor stars have kinematics suggesting 
they are different from the bulge giants, although the sample size is small ($\sim$50).  It has been put
forward that these metal-poor stars are actually halo interlopers \citep[e.g.,][]{howes14,
kunder15, koch16}.}
\label{N6441_rot_curve}
\end{figure*}

Figure~\ref{N6441_rot_curve} shows the bulge/bar rotation curve traced out by the bulge 
giants from the Bulge Radial Velocity Assay (BRAVA) survey \citep{kunder12} compared 
with both the metal-poor field bulge stars and the field RRLs surrounding NGC~6441.  
The average velocities and dispersions of $\sim$100 stars with $\rm [Fe/H] < -$1~dex from the 
ARGOS survey \citep{ness13} are also over-plotted.  It is not clear how metal-poor these stars 
reach or where exactly they are located because we do not have access to individual measurements; 
unlike the bulge surveys mentioned previously, the ARGOS data is not-public.  
The field RRLs around NGC~6441 are slower-rotating compared with the giants, as seen
also for the inner Galaxy RRLs in \citet{kunder16}.  However, this field extends further in 
longitude than probed in \citet{kunder16}.  A more extensive investigation
of the bulge field RRLs covering a larger spatial across the bulge is being carried out and will 
be the subject of a separate paper (Kunder et al., in prep).

\subsection{Orbit}
\label{orbit}
Material lost by accreted dwarf galaxies is predicted to form pronounced tidal tails on both sides of 
the dwarf: the leading tail, traveling faster that the dwarf, and the trailing tail which moves slower,
and for circular orbits (like NGC~6441) the tails are tracers of the cluster
path \citep[e.g.,][]{jordi10, kazantzidis11, lokas11}.
The strong tidal tails of the globular clusters, Palomar~5 and NGC~5466, for example, roughly follow the 
orbit of the cluster \citep{belokurov06, grillmair06} with some (expected) mis-alignment.  

To gain insight into where expected tidal debris would lie, an orbital integration of NGC~6441 is
carried out.  We adopt the radial velocity of the cluster presented here, along with its 
distance of 11.6~kpc from \citet[from][2010 edition]{harris96} and proper motion from 
\citet{casetti10}.
Using the {\tt galpy} Python package\footnote{http://github.com/jobovy/galpy; version 1.2} 
and the recommended {\tt MWPotential2014} potential 
with the default parameters \citep{bovy15}, we integrated the orbit in time for 150 Myr. 
The orbital path is shown by the dotted line in Figure~\ref{levi_mathew} (left panel).  
The correlation between the orbit of NGC~6441 orbit and the candidate extra-tidal stars is suggestive 
and indicates a low probability that the 
all of these stars are due to random fluctuations in the field.  

Calculating the orbit of NGC~6441 also gives information as to where this cluster 
may have originated.  As we have  
integrated forward in time by 150 Myr, the trailing arm will lie almost directly behind the leading arm.
If the orbital path of NGC~6441 is tracing out the leading arm, most of our 3$\sigma$ candidates are 
located along the trailing arm, as is expected for tidal debris.  
From observations of known streams in the Milky Way, ``narrow" streams have widths of
$\sim$0.25 degrees \citep{koposov14} and prominent streams stemming from globular clusters (like Pal 5) have
widths of $\sim$0.75 degrees.  NGC~6441 is more massive than most GCs and also has a relatively large 
radial velocity dispersion, so it would not be surprising if the tidal debris has a width of at least $\sim$0.75
along its orbit.  Careful $N$-body modeling of this cluster could yield more details as to the expected 
width and density of tidally stripped stars. 


\section{Discussion}
RRLs are not nearly as ubiquitous throughout the bulge as compared to bulge red clump stars 
and red giants.  They are therefore much less likely to be field star contaminants if they are located
around stellar systems, such as globular clusters, with rich populations of RRLs. 
RRLs have also been shown to be particularly useful in finding streams in the inner Galaxy, even with the lack 
of kinematical information, shedding important detections on dissolving GCs in the inner Galaxy \citep{mateu17}. 

Here we present radial velocities of 40 RRLs around the massive, complex and peculiar GC NGC~6441
to investigate this clusters spatial extent and its connections to the bulge field RRLs.  In particular,
we try to find candidate dissolved cluster members stemming directly from NGC~6441, a cluster 
in which the RRLs are a second of third generation population \citep{lee14}, similar to the 
field stars in the inner Galaxy with abundances that are typically found in GC stars presented 
in \citet{schiavon17a, fernandeztrincado17}

As was also seen by \citet{gratton07} who identified NGC~6441 members from high-resolution
abundances, we find evidence that the tidal radius of NGC~6441 is underestimated by a factor of at 
least two (Figure~2, right panel).  We also identify RRL candidates which may
be part of tidal tails originating from the cluster, or horizontal branch stars being dynamically 
evaporated \citep[e.g.,][]{krogsrud13}, or part of a large diffuse stellar envelope, such
as seen surrounding the halo GCs NGC~1851, NGC~5824, M2, and NGC~1261 
\citep{olszewski09, kuzma16, kuzma18}.  These stars have periods, amplitudes and magnitudes 
that are similar to what is seen within the bonafide NGC~6441 stars.  Proper motions of these stars 
from Gaia as well as follow-up high-resolution spectroscopy will be invaluable to confirm their membership.  
In particular, it would be of interest to see if these stars are N-rich, as is found in the population of stars
by \citet{schiavon17a}.

If it is confirmed that NGC~6441 has dissolving RRLs, this could make it similar to Terzan~5, a cluster that
was much more massive in the past and likely formed at the epoch of the Milky Way bulge formation. 
This could indicate that although most of the bulge is not ``classical" made by previous major 
mergers\citep[e.g.,][]{shen10} there could be a part of the inner Galaxy where repeated mergers at high 
redshift contributed to the build up of the bulge \citep[e.g.,][]{hopkins10}. 

The RRL field stars surrounding NGC~6441 have velocities indicating they are rotating slower than
the younger giants confined to the bulge/bar.  Their velocity dispersion is indicative of a hot population 
residing in the inner Galaxy, with properties similar to the most metal-poor stars currently observed in
the bulge.  Both the RRLs in the field of the inner Galaxy as well as those in the GC NGC~6441 suggest
a connection with an earlier epoch of bulge formation than the bulge/bar.

Lastly, we note that the frequency of RRLs in a cluster is significantly lower than the frequency of
main sequence stars in a cluster, and tidal tails of GCs are primarily formed by the lowest-mass stars 
\citep[e.g.,][]{combes99, koch04}.  If the RRLs observed here are indeed signatures of 
tidal debris around NGC~6441, then deep CMDs of the outskirts of the cluster to search for the
cluster's main-sequence would allow a more complete tracing of the extent of NGC~6441, although
proper motion cleaning would be an essential element in such a study.  

Due to the dearth of spectroscopic observations of RRLs in the inner Galaxy, identifying RR Lyrae stars in 
moving groups in the bulge is a field largely unexplored.  However, the results presented here argue that 
detailed elemental abundance patterns of the inner galaxy RRLs could establish the link 
between the old, spheroidal Galactic bulge and it's possible build up from the destruction and/or evaporation 
of globular clusters.
\\
\\
\\
Software: Galpy (v1.2; Bovy 2015)

\acknowledgements
We thank the Australian Astronomical Observatory, which have made these observations possible. 
DMN was supported by the Allan C. and Dorothy H. Davis Fellowship.
Parts of this work were supported by Sonderforschungsbereich SFB 881 "The Milky Way System"  (subproject A8) of the German Research Foundation (DFG).

\appendix
Here we provide notes on individual stars in our sample. 
\\ \\
{\it OGLE-BUL-RRLYR-03748}:  This RRL was not previously listed in the \citet{clement01} catalog as 
being a NGC~6441 member.  However, it is within the tidal radius of NGC~6441 and our 
well-phased RV curve indicates it has a velocity well within the RV distribution of the cluster.  We note that 
our RV curve appears clean, despite the $\sim$ 0.3 mag 
scatter in the OGLE optical light curve, which suggests its light curve instabilities such as the Blazhko effect.  
From its OGLE photometry, we see it also has a long period and large 
$I$-amplitude (indicative of an OoII- or OoIII-type star), and it has a de-reddened magnitude and color very similar to the
bona fide NGC~6441 RRLs. 
\\ \\
{\it OGLE-BUL-RRLYR-04288}:  This RRL was previously listed in the \citet{clement01} catalog as V67 and as
being a field star.  It has a magnitude that is brighter than the majority of the NGC~6441 stars.  
It lies just within the tidal radius of NGC~6441 and our 
RV observations indicate it has a velocity well within the RV distribution of the cluster.  We note that 
there is some scatter in our RV curve, although the OGLE $I$-band light curve appears to be rather stable,
with only small ($\sim$0.05) magnitude variations.
\\ \\
{\it OGLE-BUL-RRLYR-03934}:  This star is well within the tidal radius of the cluster, and has been studied previously by
\citet{clementini05}, who determined an $\rm [Fe/H] \sim$1~dex for this star.  Similar to the RV reported by
\citet{clementini05}, this star has a RV 2-3 $\sigma$ lower than the cluster mean.
We note that our three radial velocities have a $\sim$15 km~s$^{-1}$ scatter around the radial velocity template.  A 
consistent $\sim$0.05 mag scatter is also seen in the phased OGLE $I$-band light curve. 
\\ \\
{\it OGLE-BUL-RRLYR-03774}: This star is the closest RRL in our sample that is outside the nominal tidal radius 
of NGC~6441 ($r\sim$8 arc min).  Its radial velocity is on the high end of the velocity distribution, but 
well within 2-$\sigma$ of the cluster velocity.  It has a well-phased RV curve, and from the OGLE 
photometry, we see it has a large $I$-amplitude (indicative of an OoII- or OoIII-type star). 
\\ \\
{\it OGLE-BUL-RRLYR-03613}: This star has a well-phased RV curve, and a velocity placing it well within what the RV distribution
for NGC~6441.  However, it is outside the cluster's tidal radius ($r\sim$13 arc min).  Its pulsation properties indicate it is an OoII or OoIII-type star, as it has a long period and large $I$-amplitude. 
\\ \\
{\it OGLE-BUL-RRLYR-31080}:  This star has a well-phased RV curve, and a velocity placing it well within what the RV distribution
for NGC~6441.  However, it is well-outside the cluster's tidal radius ($r\sim$24 arc min).  It is spatially located along the leading arm
of the cluster and has a large $I$-band amplitude.
\\ \\
{\it OGLE-BUL-RRLYR-30928}:  We have two noisy measurements at $\phi$$\sim$0.0, and one clean one 
at $\phi$$\sim$0.8.  Using all three measurements, the radial velocity template fits best at a velocity 
of $\sim$ 50~km~s$^{-1}$.  Anchoring the radial velocity template to our best measurement, a velocity
of $\sim$ 30~km~s$^{-1}$ is found.  Therefore, this star could also be a $2\sigma$ cluster candidate.
It has pulsation properties indicative of an OoII or OoIII-type RRL, and is spatially located along the
leading arm of the cluster.
\\ \\
{\it OGLE-BUL-RRLYR-30792}:  This star has a velocity on the low end of the NGC~6441 RV distribution, 
but still within $3\sigma$ of the cluster velocity.  It is
spatially located on the leading end of the cluster orbit, $\sim$16 arc min away from the cluster center.
Both it's period and amplitude are consistent with an OoII or OoIII-type RRL.  It has a magnitude and color
consistent with what is expected for RRLs residing in NGC~6441.  
\\ \\
{\it OGLE-BUL-RRLYR-30603 }and {\it OGLE-BUL-RRLYR-30513}:  These two stars have RV curves 
with observations well-spaced along the full pulsation curve and following the RV template well.  Both of their RVs 
are on the low end of the NGC~6441 RV distribution, 
but within $3\sigma$ of the cluster velocity.  They are relatively close to the tidal radius of NGC~6441, with a distances of 
$\sim$11 arc min from the cluster center.
Both follow the leading arm of the clusters orbit, and have very similar periods and
small $I$-amplitudes.
\\ \\
{\it OGLE-BUL-RRLYR-03311}:  This RRL is interesting because it lies on the leading arm of NGC~6441
and relatively close the the tidal radius ($\sim$14 arc min from cluster center).
Although we have three reliable radial velocity observations for this star, two are both at $\phi \sim$0.81, 
so our spatial coverage is limited.  We find a mean velocity is $\sim$$-$21$\pm$5 km~s$^{-1}$.  It
has a large $I$-amplitude.
\\ \\
{\it OGLE-BUL-RRLYR-03708:}  This star is interesting because it has velocity close to the edge of the 
$3\sigma$ RV distribution of NGC~6441.  Unfortunately, we have two noisy measurements 
at $\phi$$\sim$0.95 and 0.15, and one clean one 
at $\phi$$\sim$0.3.  Using all three measurements, the radial velocity template fits best at a velocity 
of $\sim -9$~km~s$^{-1}$.  Anchoring the radial velocity template to our best measurement, a velocity
of $\sim -21$~km~s$^{-1}$ is found.
\\ \\
{\it OGLE-BUL-RRLYR-30811:} This star is interesting because it has a relatively long 
period ($\sim$ 0.68~d) and large amplitude ($\sim$ 0.58 mag), 
lying between the OoII and OoIII locus in the PA diagram.  However, our 
three radial velocity measurements all indicate it has a velocity of $\sim$$-$60~km~s$^{-1}$.


\clearpage
\begin{table}
\begin{scriptsize}
\centering
\caption{RR Lyrae stars in and around NGC~6441}
\label{obs}
\begin{tabular}{p{0.35in}p{0.35in}p{0.35in}p{0.5in}p{0.45in}p{0.35in}p{0.35in}p{0.35in}p{0.48in}p{0.35in}p{1.25in}} \\ \hline
OGLE-ID & Gal $l$ & Gal $b$ & HRV$_{\phi=0.38}$ (km~s$^{-1}$) & \# Epochs & $I$-amp & $<I>$ & $<V>$ & period (d) & r (arc~min) & notes \\ \hline
\hline
03962 & $-$6.4628 & $-$5.0038 & 21$\pm$5 & 1 & 0.348 & 16.176 & 16.810 & 0.86314882 & 0.57 & V115 \\ 
03994 & $-$6.4751 & $-$5.0217 & 24$\pm$5 & 2 & 0.504 & 16.356 & 17.549 & 0.74760389 & 0.77 & V41 \\ 
03918 & $-$6.4700 & $-$4.9970 & 52$\pm$15 & 2 & 0.690 & 16.463 & 17.364 & 0.69438907 & 0.78 & V57 \\ 
03934 & $-$6.4564 & $-$4.9931 & 2$\pm$10 & 3 & 0.216 & 16.314 & 17.442 & 0.86089028 & 1.30 & V66 \\ 
04003 & $-$6.4479 & $-$5.0090 & 32$\pm$8 & 3 & 0.386 & 16.375 & 17.546 & 0.77307182 & 1.33 & V43 \\ 
04057 & $-$6.4363 & $-$5.0254 & 25$\pm$5 & 3 & 0.462 & 16.451 & 17.677 & 0.83292280 & 2.22 & V39 \\ 
03991 & $-$6.4346 & $-$4.9974 & 28$\pm$10 & 2 & 0.743 & 16.529 & 17.562 & 0.61338283 & 2.25 & V37 \\ 
03813 & $-$6.5074 & $-$4.9822 & 40$\pm$5 & 3 & 0.313 & 16.320 & 17.514 & 0.90449519 & 2.80 & V46 \\ 
03967 & $-$6.4289 & $-$4.9853 & 13$\pm$8 & 3 & 0.693 & 16.388 & 17.509 & 0.73459734 & 2.89 & V38 \\ 
04118 & $-$6.4462 & $-$5.0545 & 31$\pm$5 & 3 & 0.242 & 16.359 & 17.434 & 0.56118119 & 3.02 & V69 -- $-$RRL of $c$ type \\ 
30154 & $-$6.5195 & $-$4.9916 & 13$\pm$10 & 3 & 0.110 & 16.477 & 17.705 & 0.86016577 & 3.17 & V96 \\ 
03748 & $-$6.4500 & $-$4.9339 & 22$\pm$8 & 3 & 0.500 & 16.488 & 17.624 & 0.76976119 & 4.72 & probable cluster member \\ 
04288 & $-$6.4262 & $-$5.1010 & 20$\pm$15 & 3 & 0.558 & 15.869 & 17.044 & 0.65365337 & 6.06 & V67 \\ 
\hline
& & & & end \  \  of & cluster & tidal & radius  \\
\hline
03774 & $-$6.6015 & $-$5.0286 & 40$\pm$5 & 3 & 0.784 & 15.344 & 16.494 & 0.48114906 & 7.97 & probable cluster member/ extra-tidal candidate \\ 
30513 & $-$6.3796 & $-$5.1298 & $-$3$\pm$5 & 3 & 0.265 & 15.323 & 16.606 & 0.62117217 & 9.00 & 3$\sigma$ extra-tidal candidate \\ 
03517 & $-$6.5836 & $-$4.9090 & $-$127$\pm$5 & 2 & 0.589 & 15.891 & 17.010 & 0.45653044 & 9.12 &  \\ 
03873 & $-$6.3499 & $-$4.9112 & $-$68$\pm$8 & 3 & 0.366 & 15.882 & 17.096 & 0.61596058 & 9.33 & \\ 
03820 & $-$6.6284 & $-$5.0576 & 117$\pm$5 & 3 & 0.729 & 15.349 & 16.425 & 0.52560351 & 9.92 &  \\ 
30603 & $-$6.3425 & $-$5.1531 & $-$3$\pm$5 & 3 & 0.256 & 16.065 & 17.274 & 0.60905742 & 11.50 &  3$\sigma$ extra-tidal candidate \\ 
03443 & $-$6.6478 & $-$4.9159 & 115$\pm$10 & 2 & 0.597 & 15.874 & 17.021 & 0.55429983 & 12.07 &  \\ 
03613 & $-$6.6856 & $-$5.0129 & 31$\pm$5 & 3 & 0.615 & 14.552 & 15.781 & 0.62322311 & 12.94 & extra-tidal candidate \\ 
03328 & $-$6.6199 & $-$4.8478 & 103$\pm$5 & 2 & 0.601 & 15.704 & 16.885 & 0.51843990 & 13.25 &  \\ 
03311 & $-$6.5605 & $-$4.8036 & $-$21$\pm$5 & 3 & 0.736 & 15.629 & 16.928 & 0.49660290 & 13.52 &  \\ 
03708 & $-$6.6981 & $-$5.0645 & $-$9$\pm$10 & 3 & 0.820 & 16.160 & 17.378 & 0.44560223 & 14.07 &  \\ 
04306 & $-$6.2304 & $-$4.9915 & $-$30$\pm$5 & 3 & 0.720 & 16.360 & 17.670 & 0.61515817 & 14.42 &  \\ 
30563 & $-$6.2375 & $-$5.0764 & $-$110$\pm$10 & 2 & 0.401 & 15.681 & 16.883 & 0.61219646 & 14.51 &  \\ 
03380 & $-$6.4175 & $-$4.7508 & $-$207$\pm$5 & 2 & 0.817 & 15.730 & 17.104 & 0.56729551 & 15.87 &  \\ 
04049 & $-$6.6832 & $-$5.1670 & $-$120$\pm$8 & 3 & 0.763 & 15.633 & 16.881 & 0.49413515 & 15.87 &  \\ 
30792 & $-$6.3626 & $-$5.2613 & $-$5$\pm$10 & 3 & 0.494 & 16.663 & 17.752 & 0.68498653 & 16.40 &  3$\sigma$ extra-tidal candidate \\ 
03462 & $-$6.3510 & $-$4.7478 & $-$160$\pm$5 & 2 & 0.488 & 15.992 & 17.198 & 0.57118988 & 17.28 &  \\ 
03168 & $-$6.5852 & $-$4.7455 & 25$\pm$15 & 3 & 0.329 & 15.835 & 17.046 & 0.70627368 & 17.31 & extra-tidal candidate \\ 
30651 & $-$6.5478 & $-$5.2925 & 360$\pm$5 & 2 & 0.666 & 15.969 & 17.108 & 0.53130980 & 17.58 &  \\ 
30926 & $-$6.3680 & $-$5.3233 & $-$146$\pm$5 & 2 & 0.674 & 16.035 & 17.264 & 0.51673355 & 19.77 &  \\ 
30928 & $-$6.2351 & $-$5.2472 & 50$\pm$10 & 3 & 0.738 & 15.425 & 16.513 & 0.64609600 & 20.03 & 3$\sigma$ extra-tidal candidate \\ 
03498 & $-$6.2676 & $-$4.7158 & $-$185$\pm$5 & 3 & 0.800 & 15.475 & 16.727 & 0.56468901 & 21.43 &  \\ 
30855 & $-$6.5029 & $-$5.3747 & $-$62$\pm$10 & 2 & 0.576 & 16.491 & 17.522 & 0.48411817 & 21.97 &  \\ 
30811 & $-$6.5678 & $-$5.3917 & $-$61$\pm$5 & 3 & 0.585 & 14.956 & 15.968 & 0.67843067 & 23.64 &  \\ 
31080 & $-$6.2743 & $-$5.3643 & 41$\pm$5 & 3 & 0.764 & 15.692 & 16.622 & 0.51081605 & 24.29 &  extra-tidal candidate \\ 
30975 & $-$6.5204 & $-$5.4417 & $-$57$\pm$5 & 3 & 0.751 & 15.858 & 16.856 & 0.48111461 & 26.08 &  \\ 
30939 & $-$6.6423 & $-$5.4911 & 85$\pm$5 & 3 & 0.521 & 15.903 & 16.988 & 0.56576299 & 30.66 &  \\ 
\end{tabular}
\end{scriptsize}
  \end{table}

\clearpage

\end{document}